\documentclass[aps,prl,floatfix,twocolumn,preprintnumbers,amsmath,amssymb,groupedaddress,showpacs,showkeys,superscriptaddress]{revtex4-1}

\usepackage{graphicx}  
\usepackage{dcolumn}   
\usepackage{bm}        
\usepackage{color}     
\usepackage{amsmath,amssymb}
\usepackage{psfrag}
\usepackage{multirow}
\usepackage{bbold}
\usepackage{braket}  

\begin{document}
\title{Quantum Anomalous Hall Effects in Graphene from Proximity-Induced Uniform and Staggered Spin-Orbit and Exchange Coupling} 
\author{Petra H\"{o}gl}
\email[Emails to: ]{petra.hoegl@physik.uni-regensburg.de}
\author{Tobias Frank}
\author{Klaus Zollner}
\author{Denis Kochan}
\affiliation{Institute for Theoretical Physics, University of Regensburg, 93040 Regensburg, Germany}
\author{Martin Gmitra}
\affiliation{Department of Theoretical Physics and Astrophysics, Pavol Jozef \v{S}af\'{a}rik University, 04001 Ko\v{s}ice, Slovakia}
\author{Jaroslav Fabian}
\affiliation{Institute for Theoretical Physics, University of Regensburg, 93040 Regensburg, Germany}

	\begin{abstract}
		\vspace{0.05cm}
		We investigate an effective model of proximity modified graphene (or symmetrylike materials) with broken time-reversal symmetry. 
		We predict the appearance of quantum anomalous Hall phases by computing bulk band gap and Chern numbers 
		for benchmark combinations of system parameters. Allowing for staggered exchange field enables quantum anomalous Hall 
		effect in flat graphene with Chern number $C=1$. We explicitly show edge states in zigzag and armchair nanoribbons and explore 
		their localization behavior. Remarkably, the combination of staggered intrinsic spin-orbit and uniform exchange coupling gives 
		topologically protected (unlike in time-reversal systems) pseudohelical states, whose spin is opposite in opposite zigzag edges. 
		Rotating the magnetization from out of plane to in plane makes the system trivial, allowing to control topological phase 
		transitions. We also propose, using density functional theory, a material platform---graphene on Ising antiferromagnet
		MnPSe$_3$---to realize staggered exchange (pseudospin Zeeman) coupling.
	\end{abstract}

	\maketitle
	
	Topological effects in graphene~\cite{Ren2016} attract immense attention due to their
	fascinating physics and potential applications in dissipationless electronics and spintronics~\cite{Zutic2004, Han2014}. 
	In the quantum spin Hall effect topological edge states are protected by time-reversal symmetry
	\cite{Haldane1988, Kane2005,Kane2005a,Qiao2010, Frank2018}, while the presence
	of an exchange coupling in the quantum anomalous Hall effect (QAHE) breaks time-reversal symmetry, inducing topological protection described by Chern numbers 
	\cite{Qiao2010,Qiao2012,Zhang2015,Zhang2015a,Zanolli2018,Zhang2018,Su2017}.
	
	There has been enormous experimental progress {\it towards}
	realizing topological phases in graphene, by means of van der Waals
	heterostructures~\cite{Geim2013}. In pristine graphene 
	intrinsic spin-orbit coupling (SOC) is predicted to be weak, 12 $\mu\mathrm{eV}$~\cite{Gmitra2009};
	a recent 
	experiment for graphene on SiO$_2$ found it to be 20 
	$\mu\mathrm{eV}$~\cite{Sichau2019:PRL}.
	However, graphene on transition-metal dichalcogenides (TMDCs), such as MoS$_2$ or WSe$_2$, 
	exhibits {\it proximity} SOC on meV scale 
	\cite{Gmitra2015,Gmitra2016,Wang2015,Yang2016,Wang2016,Volkl2017,Avsar2014,Omar2017,Dankert2017,Offidani2017,Kaloni2014,Zihlmann2018}. 
	But not only the magnitude of SOC becomes giant (compared to pristine graphene),
	the functional form of SOC changes as well. Instead of intrinsic SOC, graphene on TMDCs
	acquires staggered (valley-Zeeman) SOC with opposite sign on $A$ and $B$ sublattices, as
	spectacularly confirmed by spin relaxation anisotropy
	experiments~\cite{Ghiasi2017, Cummings2017, Benitez2018}. Staggered SOC leads to protected
	pseudohelical edge states~\cite{Frank2018} in the absence of magnetic interactions. 
	
	Adding magnetic exchange breaks time-reversal symmetry and together with SOC leads to QAHE. What if
	exchange coupling were also staggered, realizing proximity antiferromagnetic graphene? Thus far 
	only uniform (ferromagnetic) exchange coupling was considered in the proximity effect of graphene~\cite{Zollner2016,Predrag2016, Wang2015a,Leutenantsmeyer2017,Mendes2015,Swartz2012,Wei2016,Haugen2008,Yang2013,Hallal2017,Dyrdal2017,Zollner2018}. Fortunately, there are now suitable
	candidates, layered semiconducting Ising antiferromagnets, which could serve as substrate for graphene
	and induce staggered exchange. One example is discussed below. 
	\begin{figure}[]
		{\includegraphics[width=1\columnwidth]{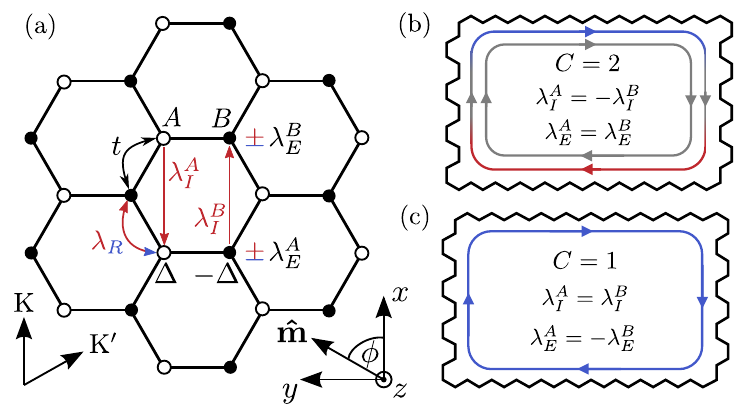}}
		\caption{(a) Scheme of graphene lattice with proximity induced hoppings. Sublattices $A$ and $B$ 
			are denoted by empty and full dots, respectively. Color indicates action on spin. The minimal model contains spin neutral 
			nearest-neighbor hopping $t$ and on site staggered potential $\Delta$; spin-mixing nearest-neighbor 
			Rashba SOC $\lambda_R$; spin and sublattice resolved next nearest-neighbor intrinsic 
			SOC $\lambda^A_I$, $\lambda^B_I$; on site sublattice resolved exchange 
			splitting $\lambda^A_E$, $\lambda^B_E$ (spin-dependent energy shift spin up $+$, spin down 
			$-$). Orientation of reciprocal lattice is shown by $\mathrm{K}$ and $\mathrm{K}'$. Magnetization orientation in 
			real space is specified by $\mathbf{\hat{m}}$. Sketch of quantum anomalous Hall states along zigzag and armchair 
			edges for (b) $\lambda^A_I=-\lambda^B_I$ and $\lambda^A_E=\lambda^B_E$ and (c) $\lambda^A_I=\lambda^B_I$ and 
			$\lambda^A_E=-\lambda^B_E$. The Chern number is given by $C$. Color indicates $\hat{s}_z$ spin expectation value: 
			red spin up, blue spin down, gray unpolarized. }
		\label{fig:1}
	\end{figure}
	
	Our main goal is to make specific predictions for topological phases in graphene (or symmetrylike materials) by considering those newly unveiled staggered
	regimes for spin-orbit and exchange couplings. In addition to QAHE phases with 
	Chern number $C=2$ (as in Kane-Mele models for uniform couplings), we find $C=1$ (single edge state) for uniform SOC and staggered exchange.
	Remarkably, rotating the exchange to the plane makes the system topologically trivial, making the magnetoanisotropy an effective knob on the 
	topological properties. 
	To support our model calculations we also introduce a platform for
	making graphene a proximity antiferromagnet. Using density functional theory, we
	calculate the electronic states of graphene on monolayer MnPSe$_3$, which is an Ising-type antiferromagnetic semiconductor. Graphene's 
	Dirac states are well preserved displaying clear signatures of staggered (pseudospin Zeeman) exchange, also confirmed
	by tight-binding fitting.

	\paragraph{Model.}
	We consider graphene modified by a proximity effect such that the sublattice, horizontal reflection, and time-reversal symmetries are broken. 
	The minimal $C_{3v}$-symmetric Hamiltonian can be written as
	\begin{eqnarray}\label{eq:hamiltonian}
	\mathcal{H} &=&
	-t\sum_{\left<i,j\right>,\sigma} c_{i\sigma}^\dagger c^{\phantom\dagger}_{j\sigma}+
	\Delta\sum_{i,\sigma} \, \xi_i \,c_{i\sigma}^\dagger c^{\phantom\dagger}_{i\sigma} \nonumber \\
	& &+\frac{2i\lambda_R}{3}\sum_{\left<i,j\right>,\sigma,\sigma^\prime}c_{i\sigma}^\dagger 
	c^{\phantom\dagger}_{j\sigma^\prime}\left[\left(\hat{\mathbf{s}}\times 
	\mathbf{d}_{ij}\right)_z\right]_{\sigma\sigma^\prime}\nonumber\\
	& &+\frac{i}{3\sqrt{3}}\sum_{\left<\left<i,j\right>\right>,\sigma,\sigma^\prime}\lambda_{ I}^{i}\nu_{ij}\,c_{i\sigma}^\dagger 
	c^{\phantom\dagger}_{j\sigma^\prime} \left[\hat{\mathbf{s}}_z \right]_{\sigma\sigma^\prime}\nonumber \\
	& &+\sum_{i,\sigma,\sigma^\prime}\lambda_E^{i}\,c_{i\sigma}^\dagger c^{\phantom\dagger}_{i\sigma^\prime} 
	\left[\mathbf{\hat{m}}\cdot\mathbf{\hat{s}}\right]_{\sigma\sigma^\prime},
	\end{eqnarray}
	where $c_{i\sigma}^\dagger\left(c_{i\sigma}^{\phantom\dagger}\right)$ is the creation (annihilation) operator for an 
	electron on lattice site $i$ that belongs to the sublattice $A$ or $B$ and carries spin $\sigma$. 
	The hoppings are depicted in Fig.~\ref{fig:1}(a). The first two terms are the spin-preserving nearest-neighbor hopping 
	(sum over $\left<i,j\right>$) and the staggered on site potential $\Delta$ with $\xi_{i}=1$ on sublattice $A$ and 
	$\xi_{i}=-1$ on $B$. The potential difference on $A$ and $B$ takes into account the different environment that 
	atoms in sublattices $A$ and $B$ encounter in a heterostructure (implicitly broken sublattice symmetry). The next 
	two terms describe SOC~\cite{Kochan2017}. 
	Rashba SOC $\lambda_R$ mixes spins of nearest neighbors, where the unit vector $\mathbf{d}_{ij}$ points from site $j$ 
	to $i$ and $\mathbf{\hat{s}}$ contains spin Pauli matrices. It occurs when inversion symmetry is broken, e.g., in an 
	asymmetric heterostructure. Intrinsic SOC couples the same spins on next-nearest neighbors (sum over 
	$\left<\left<i,j\right>\right>$). It depends on clockwise ($\nu_{ij}=-1$) or counterclockwise ($\nu_{ij}=1$) 
	hopping paths from site $j$ to $i$ and is sublattice resolved, $\lambda^i_I$ with $i=A,B$. The above described 
	terms form an experimentally relevant model for nonmagnetic graphene 
	proximity systems~\cite{Frank2018,Gmitra2013,Gmitra2016,Kochan2017} extending the earlier models from McClure and Yafet~\cite{McClure1962}, Haldane 
	\cite{Haldane1988}, and Kane and Mele~\cite{Kane2005}. The last term in 
	Eq.~(\ref{eq:hamiltonian}) extends the model to magnetic systems. It is a sublattice-dependent exchange coupling 
	that introduces spin magnetization and breaks 
	time-reversal symmetry~\cite{Zollner2016}. Similar to the intrinsic SOC we allow for different values $\lambda_E^i$ 
	with $i=A,B$ and orientation of magnetization along the unit vector 
	$\mathbf{\hat{m}}=\left(\cos\phi\sin\theta,\sin\phi\sin\theta,\cos\theta\right)$, where $\phi$ is measured with respect to the 
	$x$-axis and $\theta$ with respect to the $z$-axis in Fig.~\ref{fig:1}(a).
	
	The Hamiltonian in Eq.~(\ref{eq:hamiltonian}) breaks time-reversal, particle-hole, and chiral symmetry. Therefore it 
	belongs to class A of 2D quantum Hall systems and its topological nature can be determined by the Chern 
	number $C$~\cite{Thouless1982,Ryu2010}.
	A nonzero $C$ indicates a topological system, namely, a QAHE phase.
	As for the quantum Hall state the Chern number 
	gives the quantized Hall conductance $\sigma_H=Ce^2/h$~\cite{Thouless1982}. Every change of the Chern number is 
	accompanied by the closing of the bulk band gap.
	
	Since we wish to present material-independent topological phase diagrams, we fix the model parameters to generic 
	values $\Delta=0.1t$ and $\lambda_R=(3/2)\times 0.05t$, and explore the 
	interplay of intrinsic SOC and exchange coupling first for the out-of-plane magnetization $\mathbf{\hat{m}}=\left(0,0,1\right)$.
	For the intrinsic SOC we consider two possibilities: a {\it uniform} (u) intrinsic
	coupling, $\lambda^A_I=\lambda^B_I$, which is of the conventional
	McClure-Yafet-Kane-Mele type~\cite{McClure1962, Kane2005}, 
	recently predicted to be giant (0.5 eV) in the monolayer jacutingaite~\cite{Marazzo2018};
	a {\it staggered} (s) intrinsic SOC $\lambda^A_I=-\lambda^B_I$, which was predicted~\cite{Gmitra2015,Gmitra2016,Wang2015} and experimentally confirmed~\cite{Ghiasi2017,Benitez2018,Zihlmann2018} for graphene on TMDCs. This (also called valley-Zeeman) coupling seems 
	not to be restricted to valley-Zeeman substrates---it is also predicted to 
	come from topological insulator Bi$_2$Se$_3$~\cite{Song2018:Nano}. Similarly, we consider 
	a uniform exchange coupling, $\lambda^A_E=\lambda^B_E$, which would come from graphene on an Ising ferromagnet, and a staggered exchange,
	$\lambda^A_E=-\lambda^B_E$, which could be realized by placing 
	graphene on an Ising antiferromagnet (see below). Overall 
	we have four possible combinations: uniform-uniform (uu) $\lambda^A_I=\lambda^B_I$ and $\lambda^A_E=\lambda^B_E$, uniform-staggered (us) 
	$\lambda^A_I=\lambda^B_I$ and $\lambda^A_E=-\lambda^B_E$, staggered-uniform (su) $\lambda^A_I=-\lambda^B_I$ and $\lambda^A_E=\lambda^B_E$, 
	and staggered-staggered (ss) $\lambda^A_I=-\lambda^B_I$ and $\lambda^A_E=-\lambda^B_E$; see	Table~\ref{tab:1}.
	\begin{table}[]
		\caption{Labeling and corresponding Chern numbers for the four combinations of (out-of-plane) exchange ($E$) 
			and intrinsic ($I$) spin-orbit couplings on $A$ and $B$ atoms, summarizing Fig.~\ref{fig:2}.}
		\centering
			\scalebox{1.0}{
			\renewcommand{\arraystretch}{1.5}
			{\normalsize \begin{tabular}{l|c|c|c}
					\hline
					\hline 
					\multicolumn{2}{c|}{couplings} & $\,\lambda^A_E=\lambda^B_E \,$ & $\, \lambda^A_E=-\lambda^B_E \,$  \\
					\hline
					\multirow{2}{*}{$\lambda^A_I=\lambda^B_I$} & label 
					& (uu) & (us) \\
					& Chern & $0$,  $\pm 2$   & $0$, $\pm 1$  \\  
					\hline
					\multirow{2}{*}{$\lambda^A_I=-\lambda^B_I$ } & label 
					&   (su) & (ss) \\
					& Chern & $0$, $\pm 2$  &  $0$ \\
					\hline
					\hline
				\end{tabular}
		}}
		\centering
		\label{tab:1}
	\end{table}

\begin{figure}[b]
		\vspace{-0.4cm}
	{\includegraphics[width=1\columnwidth]{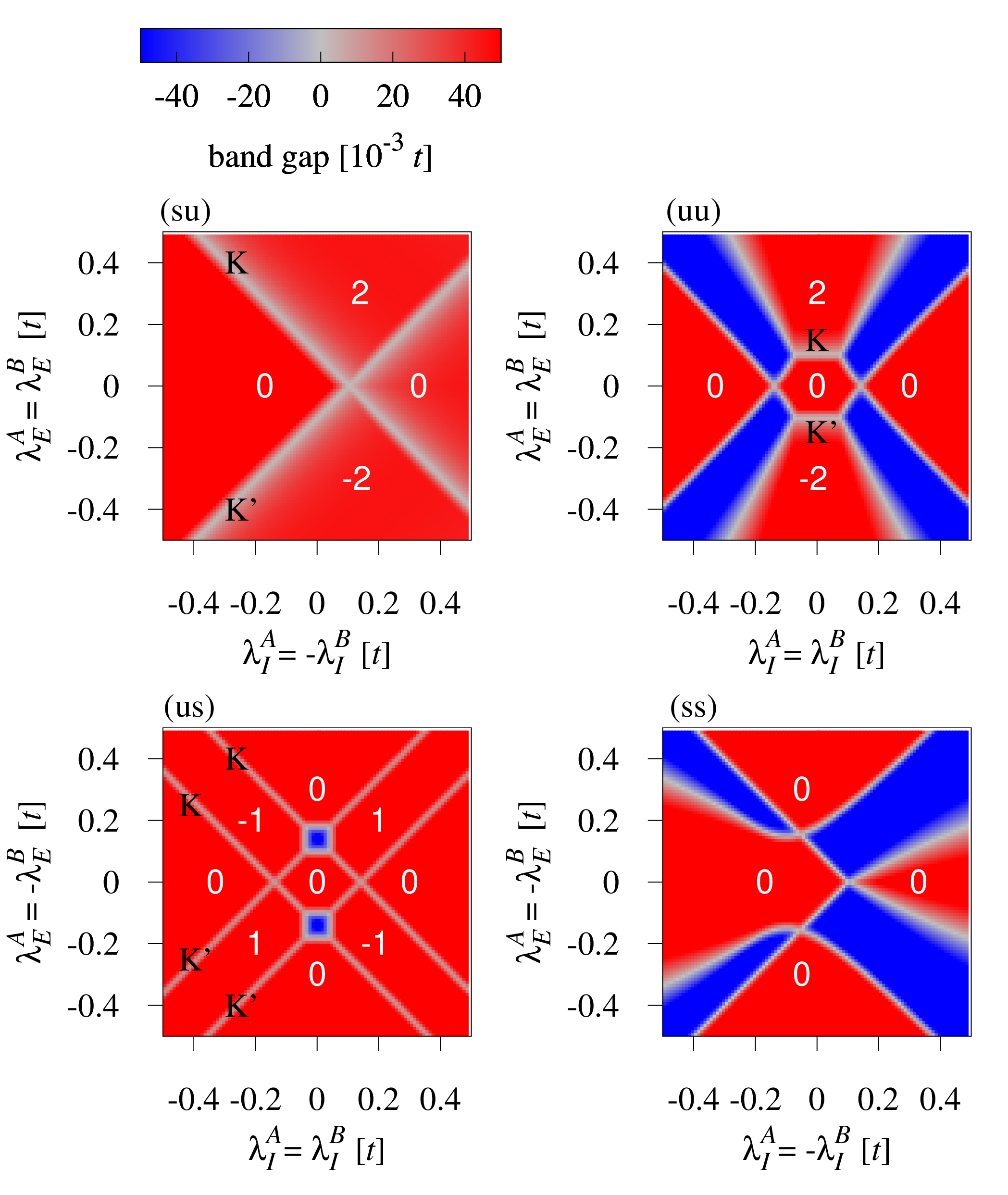}}
	\vspace{-0.6cm}
	\caption{Topological phases of magnetic graphene with SOC. Global bulk band gap and Chern 
		number (white numbers) for $\Delta=0.1t$ and 
		$\lambda_R=(3/2)\times 0.05t$ varying intrinsic SOC $\lambda^A_I$, $\lambda^B_I$ and exchange 
		splitting $\lambda^A_E$, $\lambda^B_E$ for out-of-plane magnetization 
		$\mathbf{\hat{m}}=\left(0,0,1\right)$. Position of gap closing in $\mathbf{k}$-space is 
		specified by $\mathrm{K}$, $\mathrm{K}'$. Negative band gap indicates transition to metallic system due to 
		indirect band gap closing from bands at different $\mathbf{k}$ values.}
	\label{fig:2}
	\end{figure}

	\paragraph{QAHE phases.}
	In Fig.~\ref{fig:2} we show the bulk band gap and Chern number. We find the following for the studied case: (su) two magnetic 
	Chern insulator phases with $C=\pm2$; (us) four nontrivial regions with $C=\pm1$; (uu) two QAHE phases with 
	$C=\pm2$; (ss) trivial insulating or metallic system. Analytical conditions to be in a 
	nontrivial phase are given in~\cite{supp}. Case (uu) is an extension of Ref.~\cite{Qiao2012} including both staggered 
	potential and intrinsic SOC. Cases (su) and (uu) require finite Rashba SOC to have a gapped 
	system in the vicinity of the Dirac points. For uniform SOC but staggered exchange (us), no Rashba coupling is 
	needed to induce QAHE. This case allows for single propagating states, which would be a signature for the 
	antiferromagnetic QAHE under the assumption that the absolute value of the parameters of the $A$ and $B$ sublattices do not differ drastically. 
	The discussed topological phases also exist when the exact uniform or staggered condition is relaxed as shown in the Supplemental Material~\cite{supp}. The Chern numbers are summarized in Table~\ref{tab:1}. Some aspects of the (us) and (ss) physics, 
	without staggered $\Delta$, was very recently discussed in Ref.~\cite{Luo2019:PRB}.

	\paragraph{Proposal for proximity antiferromagnetic graphene.}
	We now present a specific proposal for making graphene antiferromagnetic and introduce a material 
	platform for cases (us) and (ss), or simply
	staggered exchange. The platform is graphene on monolayer MnPSe$_3$, which is an Ising antiferromagnet 
	and semiconductor~\cite{Pei2018:FP,Wiedenmann1981:SSC}. Since bulk 
	MnPSe$_3$ is a layered compound (with optical gap 2.3 eV~\cite{Grasso1999:OSA}), 
	and only the top monolayer is important for proximity effects, the platform can be experimentally realized 
	using a MnPSe$_3$ film. There is an earlier proposal~\cite{Qiao2014} to place graphene on an antiferromagnet, perovskite BiFeO$_3$, 
	but the Fe (111) plane that proximitizes graphene is ferromagnetic, inducing a ferromagnetic (as also in~\cite{Zhang2015a,Zhang2018})
	and not antiferromagnetic exchange in graphene. 
		
\begin{figure}[b]
		\includegraphics[width=0.99\columnwidth]{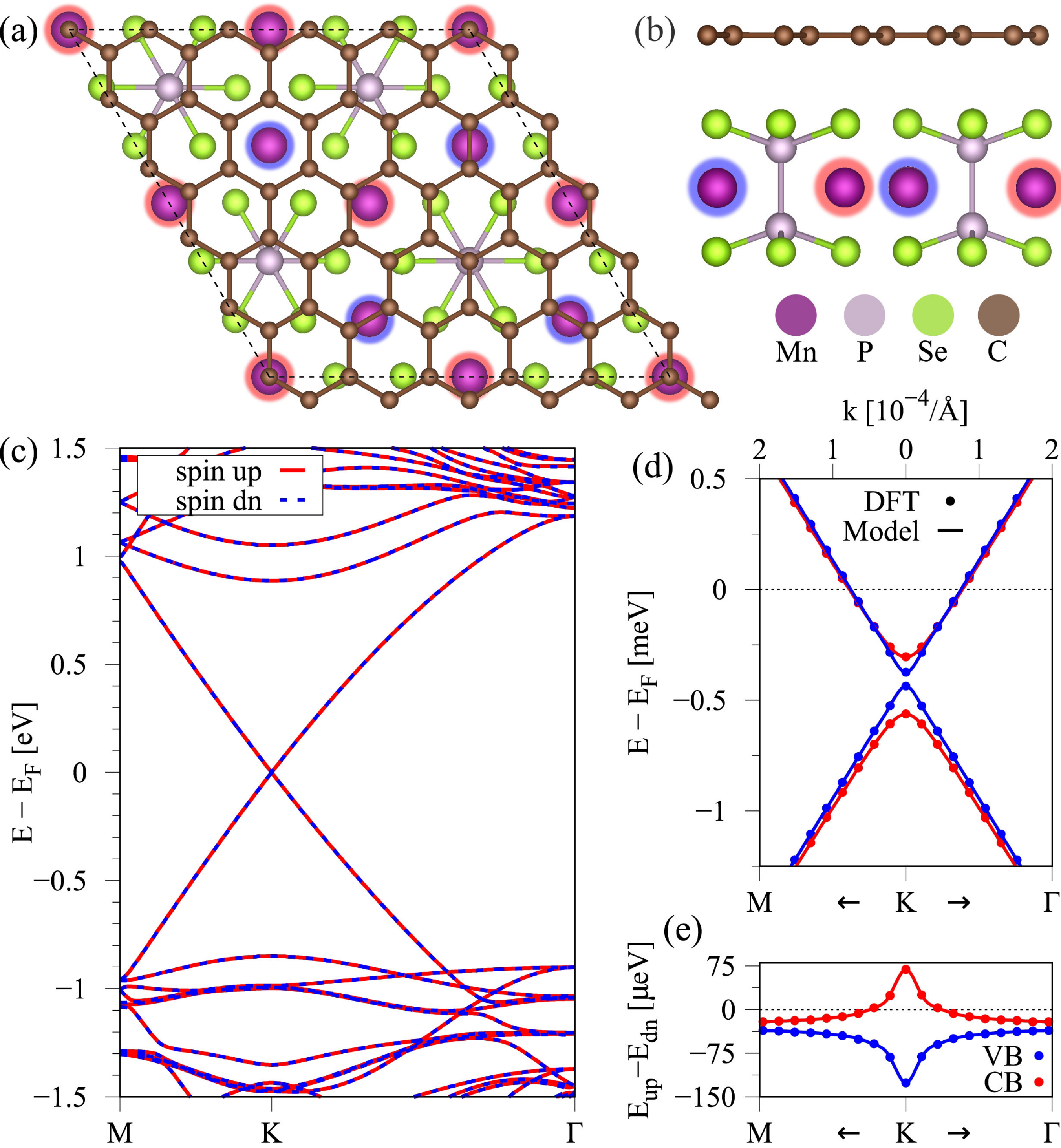}
		\caption{(a,b) Top and side view of graphene/MnPSe$_3$ heterostructure. The colored spheres
			around Mn atoms indicate the magnetization direction (red spin up, blue spin down) with Ising-type magnetic ordering. (c) Calculated electronic band structure of
			graphene/MnPSe$_3$ heterostructure without SOC (to demonstrate bare staggered exchange) 
			for an interlayer distance of $3.36$~\AA. 
			Bands in red (blue) correspond to spin up (down). (d) Enlargement of the calculated low energy bands 
			(symbols) at $\mathrm{K}$ with a fit to model Hamiltonian $\mathcal{H}_\mathrm{GR}$~\cite{supp} (solid line). 
			(e) The splitting between spin up and spin down bands $\mathrm{E}_{\textrm{up}}-\mathrm{E}_{\textrm{dn}}$ for the VB and CB.
			\label{Fig:DFT_bands_geometry}}
	\end{figure}

	In Figs.~\ref{Fig:DFT_bands_geometry}(a)-\ref{Fig:DFT_bands_geometry}(c) we show the investigated atomic and calculated electronic structure of graphene on monolayer MnPSe$_3$ where Mn forms a hexagonal lattice with 
	alternating out-of-plane magnetization. Details of the calculations are
	described in the Supplemental Material~\cite{supp}. 
	The Dirac states from graphene are nicely preserved and reside within the band gap of MnPSe$_3$. 
	An enlargement of the low energy bands around the $\mathrm{K}$ point reveals proximity-induced staggered exchange splitting of the bands.
	Our model Hamiltonian~\cite{supp} fits the low energy 
	dispersion and spin splittings, see Figs.~\ref{Fig:DFT_bands_geometry}(d,e). 
	\begin{figure*}[t!]
	{\includegraphics[width=2\columnwidth]{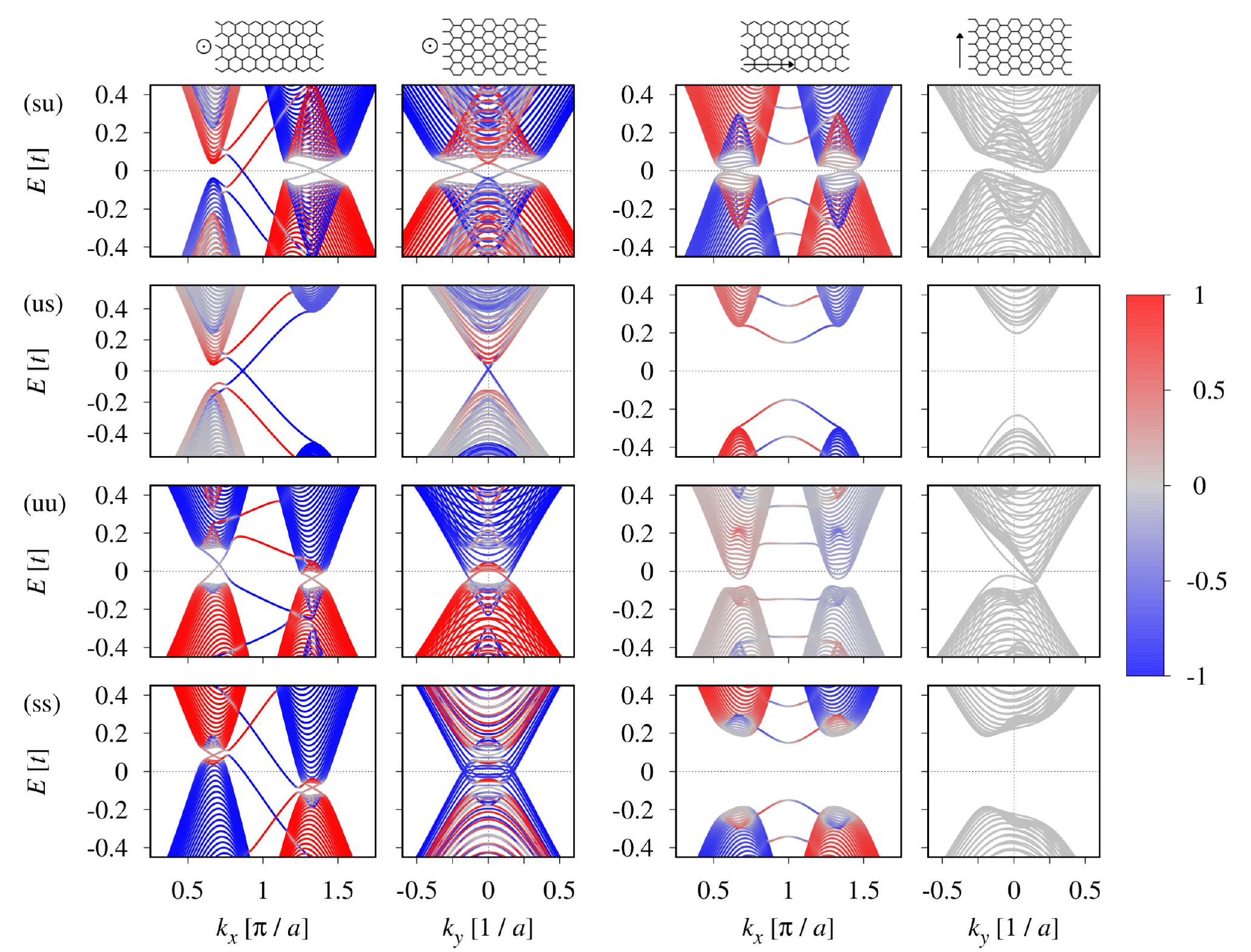}}
	\caption{Calculated electronic structure of zigzag and armchair 100 unit cells wide nanoribbons. 
		Arrows specify direction of magnetization, parallel to the $z$-axis ($\theta=0$, $\phi=0$) in the left two columns and parallel to the $x$-axis ($\theta=\pi/2$, $\phi=0$) in the right two columns. Color indicates $\hat{s}_z$ spin expectation value. We use $\Delta=0.1t$, $\lambda_R=(3/2)\times 0.05t$, $\lambda^A_E=|\lambda^B_E|=0.25t$ and $\lambda^A_I=|\lambda^B_I|=3\sqrt{3}\times 0.06t$ in 
		(su), (us), (ss), and $\lambda^A_I=|\lambda^B_I|=0.1t$ in (uu).}	
	\label{fig:4}
	\end{figure*}
	
	For our heterostructure, the splitting between spin up and spin down bands 
	$\mathrm{E}_{\textrm{up}}-\mathrm{E}_{\textrm{dn}}$ for the valence band (VB) is negative, 
	while the one for the conduction band (CB) is positive at the $\mathrm{K}$ point,
	see Fig.~\ref{Fig:DFT_bands_geometry}(e). 
	Even though the antiferromagnetic hexagonal Mn-lattice is not commensurate with the graphene lattice, we 
	effectively get different spin splittings for VB and CB.
	The fitting parameters of the low energy model Hamiltonian~\cite{supp}, 
	for several interlayer distances, show that 
	the sublattice resolved proximity exchange parameters have opposite signs.

	\paragraph{Nanoribbons.}	
	We now go back to the model investigations. Due to bulk-edge correspondence the Chern number gives the 
	number of topological states that appear per 
	edge in a nanoribbon. For zigzag and armchair termination this is displayed in the first and second column 
	of Fig.~\ref{fig:4}, respectively. The nanoribbon spectra confirm the presence of chiral edge states in the 
	QAHE phases (su), (us), and (uu)~\cite{supp}. In a finite flake the states of both termination types connect and 
	travel along the edges as schematically depicted for cases (su) and (us) in Figs.~\ref{fig:1}(b) and \ref{fig:1}(c). The (su) 
	case is curious. Without exchange, 
	this case is topologically trivial, but protected pseudohelical states emerge in a finite ribbon~\cite{Frank2018}: 
	along one zigzag edge the state has spin up, along the opposite one it has spin down. Spin flip occurs in 
	tunneling along an armchair edge. The degenerate, time-reversed partners show analog behavior. The breaking of the time-reversal symmetry makes the system topological 
	and the pseudohelical states, depicted in Fig.~\ref{fig:1}(b), become propagating also along the armchair direction. 
	Thus, in a QAHE regime, we predict protected edge states whose spin polarization is opposite in 
	opposite zigzag edges.
	
	Despite their chiral nature the QAHE states differ in their localization behavior and spin. The states' 
	position in $\mathbf{k}$-space is crucial for both. Inside a cone the edge states are much closer to bulk states and so 
	have a significantly larger decay length [e. g., case (uu) zigzag ribbon] compared to the ones that reside 
	between two cones [e.g., case (us) zigzag ribbon] (see Supplemental Material~\cite{supp}). In addition, 
	their spin polarization gets blurred inside the valleys.
	
	The zigzag ribbon in case (su) combines both types of states and allows to tune the spin-polarized intervalley modes to unpolarized 
	intravalley states by decreasing intrinsic SOC to zero. Note that for, $\lambda_I=0$, cases (su) and (uu) coincide and 
	reproduce an earlier model for QAHE states in graphene based on Rashba SOC and exchange, which has two 
	intravalley states in each valley~\cite{Qiao2010,Qiao2012}. 
	Remarkably, the intervalley modes, which are stabilized by intrinsic SOC, 
	are much more localized than the intravalley states. Therefore, the intervalley states are more robust against 
	weak disorder. But the long localization length of the intravalley states offers the possibility to tune the 
	case (su) by finite size effects. While the probability amplitude of intervalley states falls off to zero within a
	few lattice sites, the wave function of intravalley states extends over more than 30 sites (see Supplemental Material~\cite{supp}). In a small ribbon the 
	latter will hybridize and be gapped out so that we are left with only one state per edge. Similar scenario occurs 
	in the pseudohelical regime at quantum spin Hall effect~\cite{Frank2018}.
	
	In the armchair ribbon $\mathrm{K}$ and $\mathrm{K}'$ points are folded back to $\Gamma$ point, thus all edge states are within the 
	valleys and are less localized compared to zigzag edges due to the vicinity to the bulk spectrum~\cite{supp}.

	\paragraph{Magnetic anisotropy of QAHE.}
	Finally, we demonstrate that the topological order and edge states can be controlled by magnetization orientation. 
	We direct the exchange to point in plane along $x$; for uniform
	exchange this geometry was considered in Ref.~\onlinecite{Ren2016a}.
	It is clear from the presented results in Fig.~\ref{fig:4} (third and fourth columns), that \textit{in all four 
	cases the system is in a trivial phase after the rotation.} This result is elaborated
	on in the Supplemental Material~\cite{supp}, where we identify trivial 
	insulator or metallic regions in topological diagrams, and 
	also show the evolution of bulk and ribbon bands under rotation of 
	magnetization. When magnetization is induced to graphene via a small external magnetic field, as is possible in cases (su) and (uu), 
	rotating the magnetic field can be used to perform a topological phase transition where all other system parameters remain the same.
	
	In summary, a comprehensive parameter space analysis of a realistic effective model with combinations of uniform and 
	staggered spin-orbit and exchange proximitized graphene is presented, predicting magnetization orientation dependent 
	QAHE topological phases and edge states. A specific
	material platform for staggered exchange in graphene is proposed, based on DFT studies of graphene on Ising 
	antiferromagnetic MnPSe$_3$.

This work has been funded by the Deutsche Forschungsgemeinschaft (DFG, German Research Foundation) – Project-No. 314695032 – SFB 1277, DFG SPP 1666, EU Seventh Framework Programme under Grant Agreements No. 604391 Graphene Flagship, No. VVGS-2018-1227, and No. VEGA 1/0105/20.
	
\bibliography{paper}
\pagestyle{empty}	
	\begin{figure*}[htb]
		{\includegraphics[width=2.1\columnwidth]{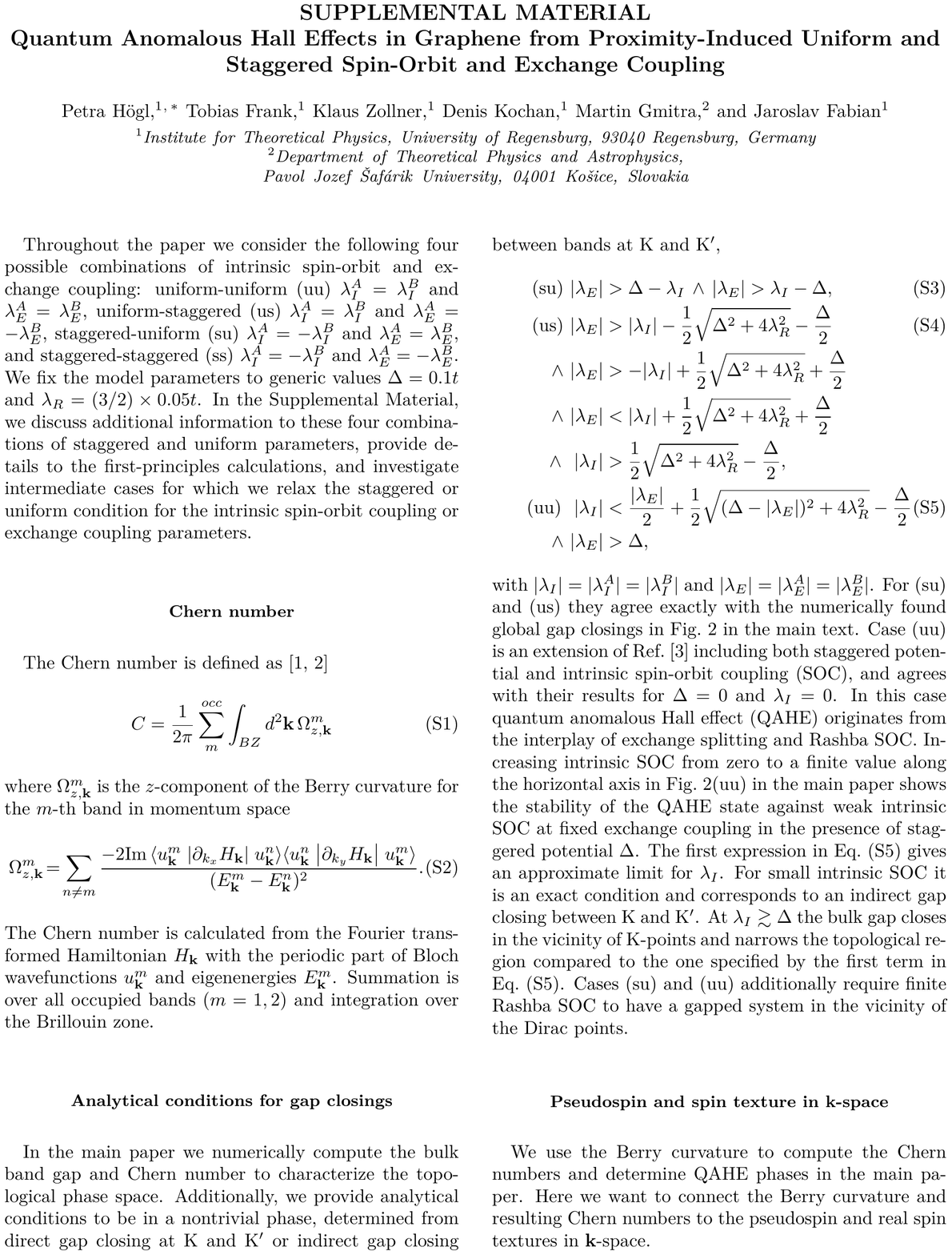}}
	\end{figure*}
	\begin{figure*}[htb]
		{\includegraphics[width=2.1\columnwidth]{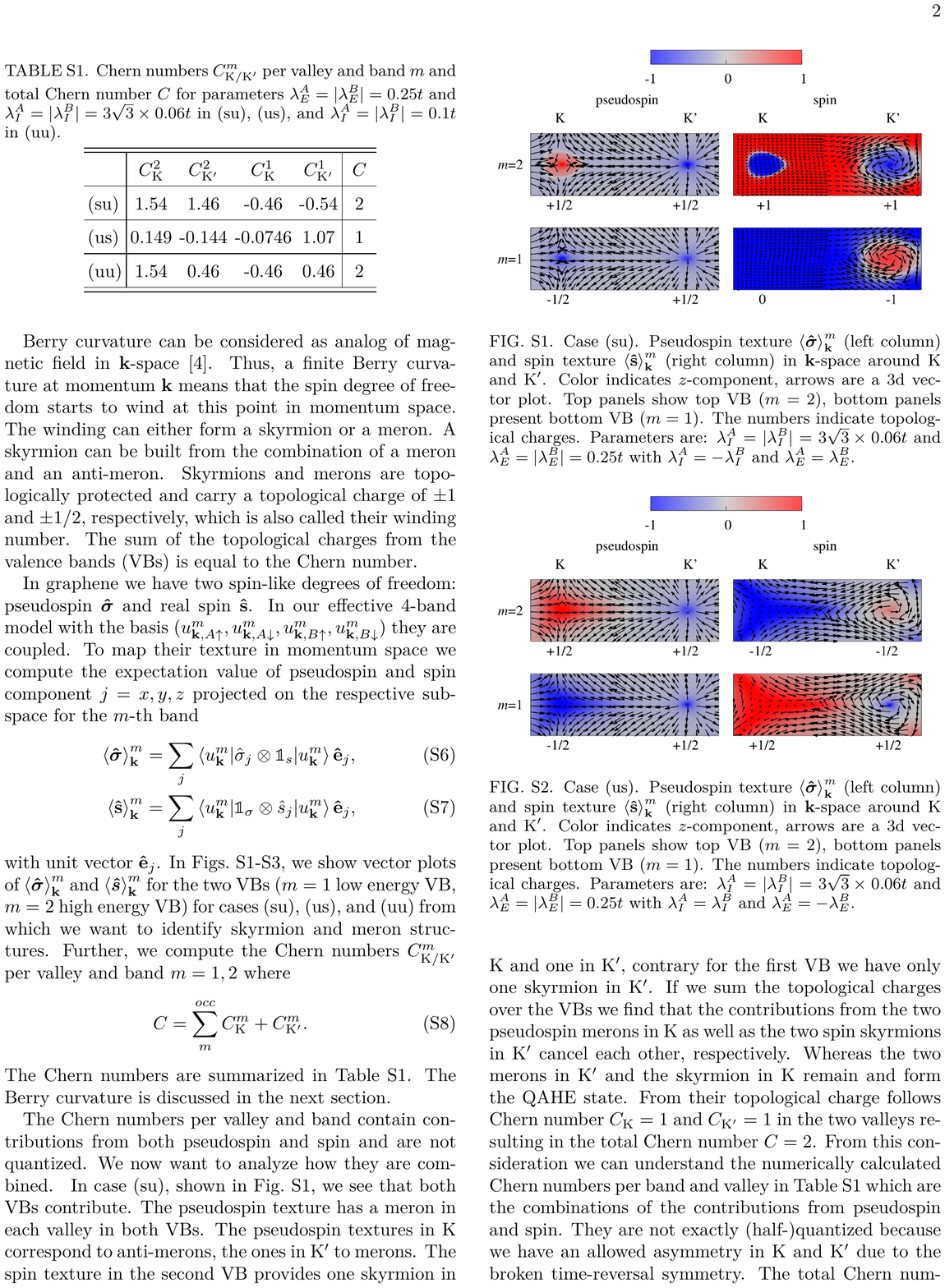}}
	\end{figure*}
\begin{figure*}[htb]
	{\includegraphics[width=2.1\columnwidth]{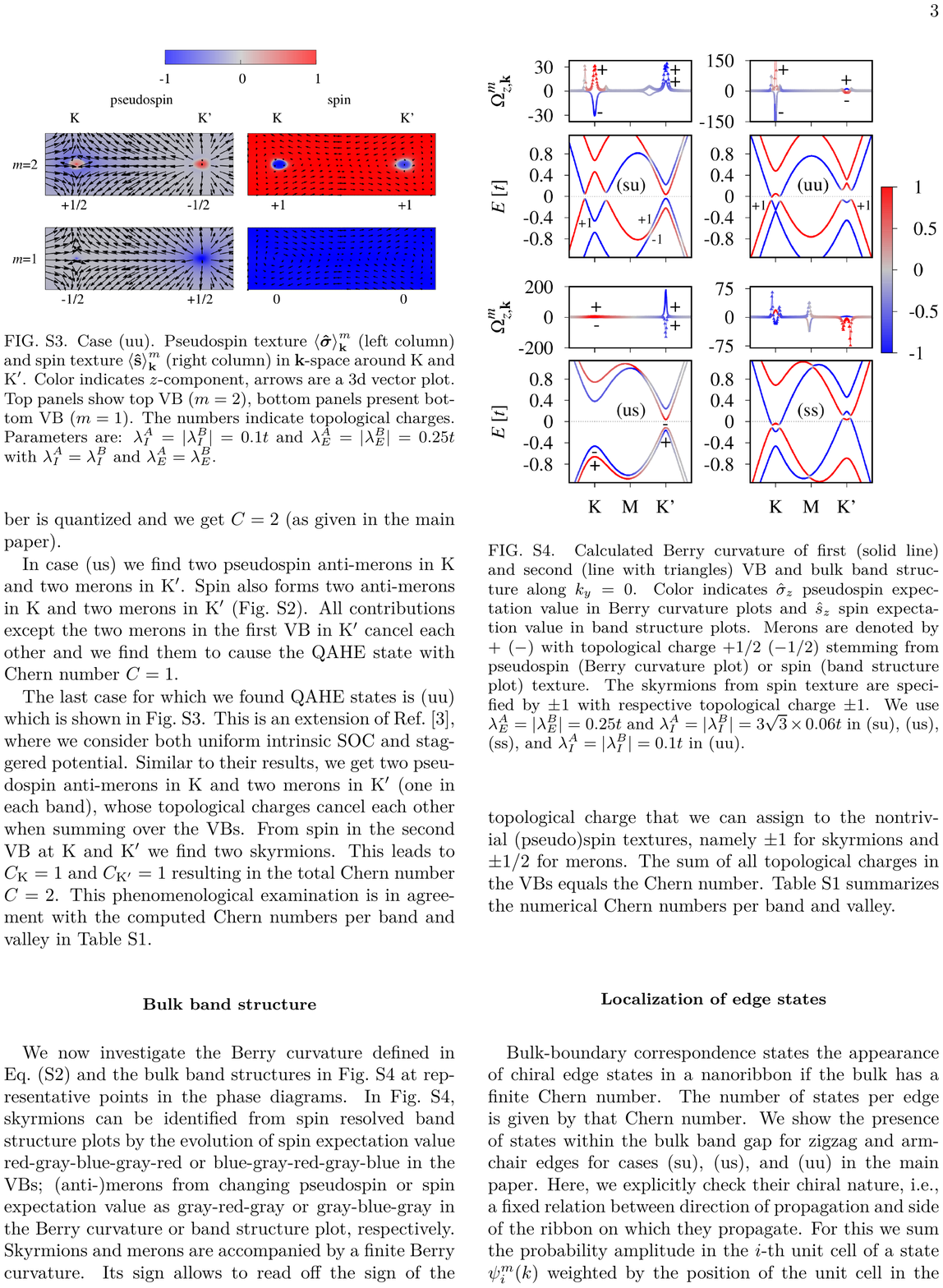}}
\end{figure*}
\begin{figure*}[htb]
	{\includegraphics[width=2.1\columnwidth]{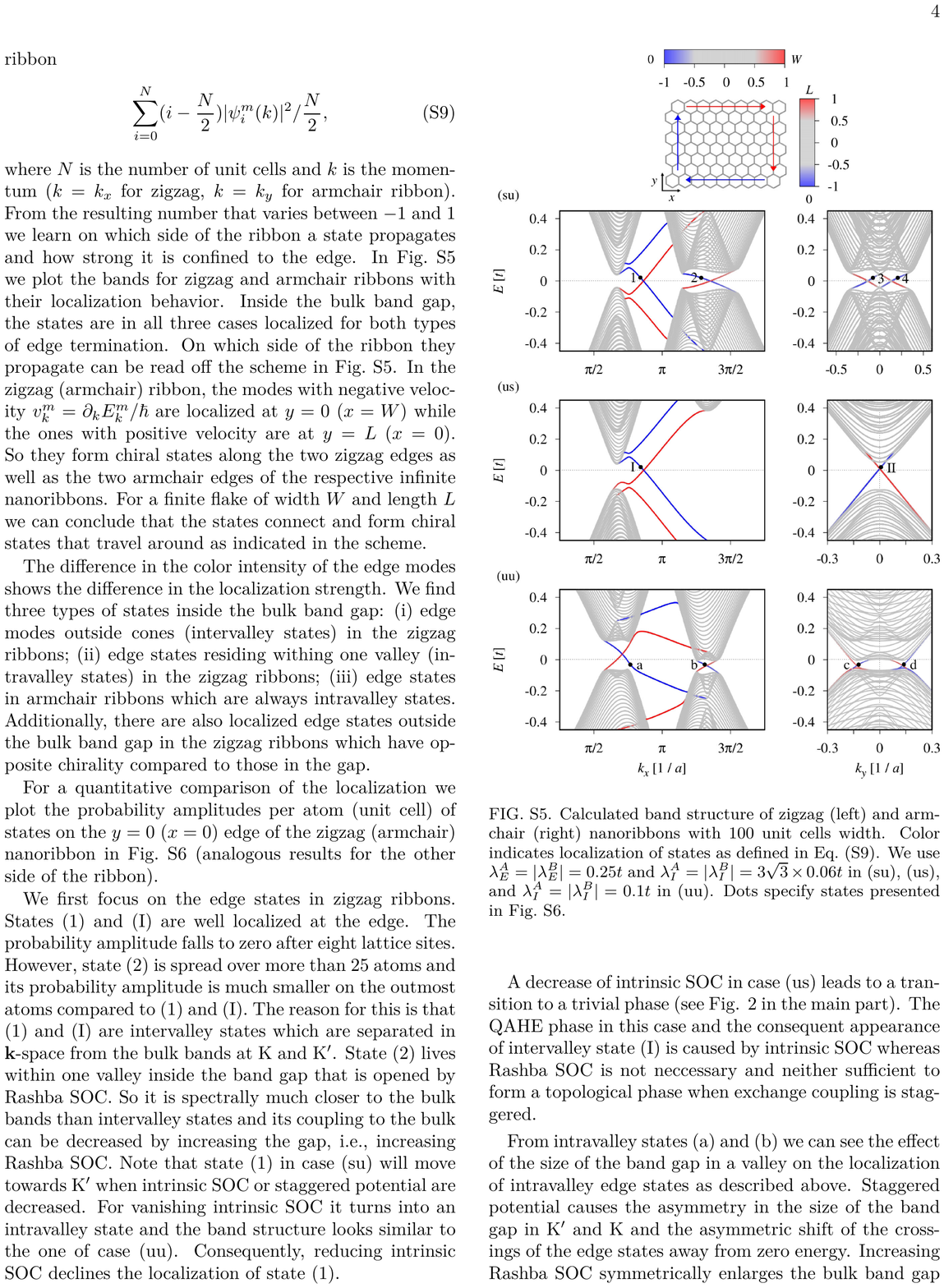}}
\end{figure*}
\begin{figure*}[htb]
	{\includegraphics[width=2.1\columnwidth]{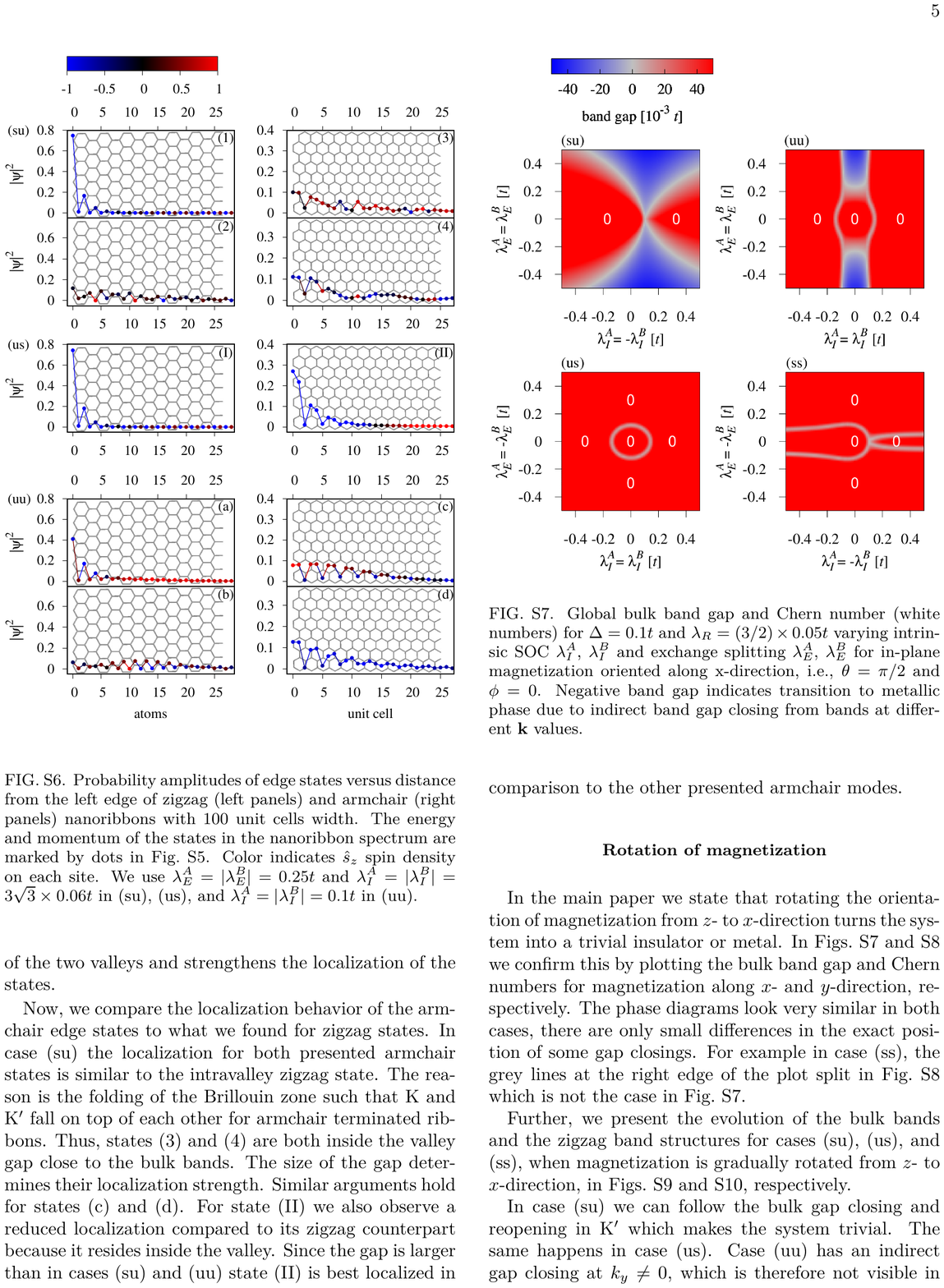}}
\end{figure*}
\begin{figure*}[htb]
	{\includegraphics[width=2.1\columnwidth]{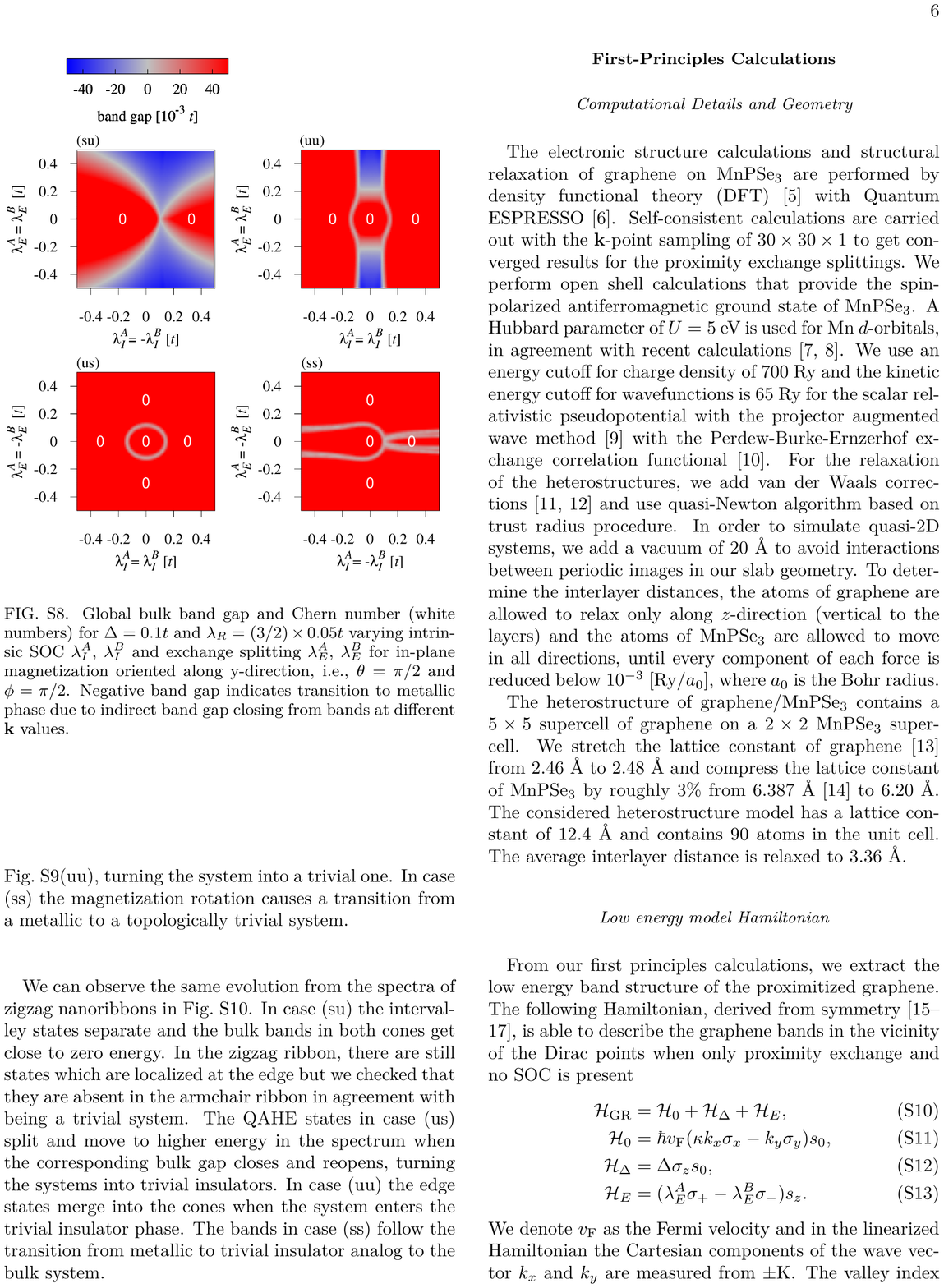}}
\end{figure*}
\begin{figure*}[htb]
	{\includegraphics[width=2.1\columnwidth]{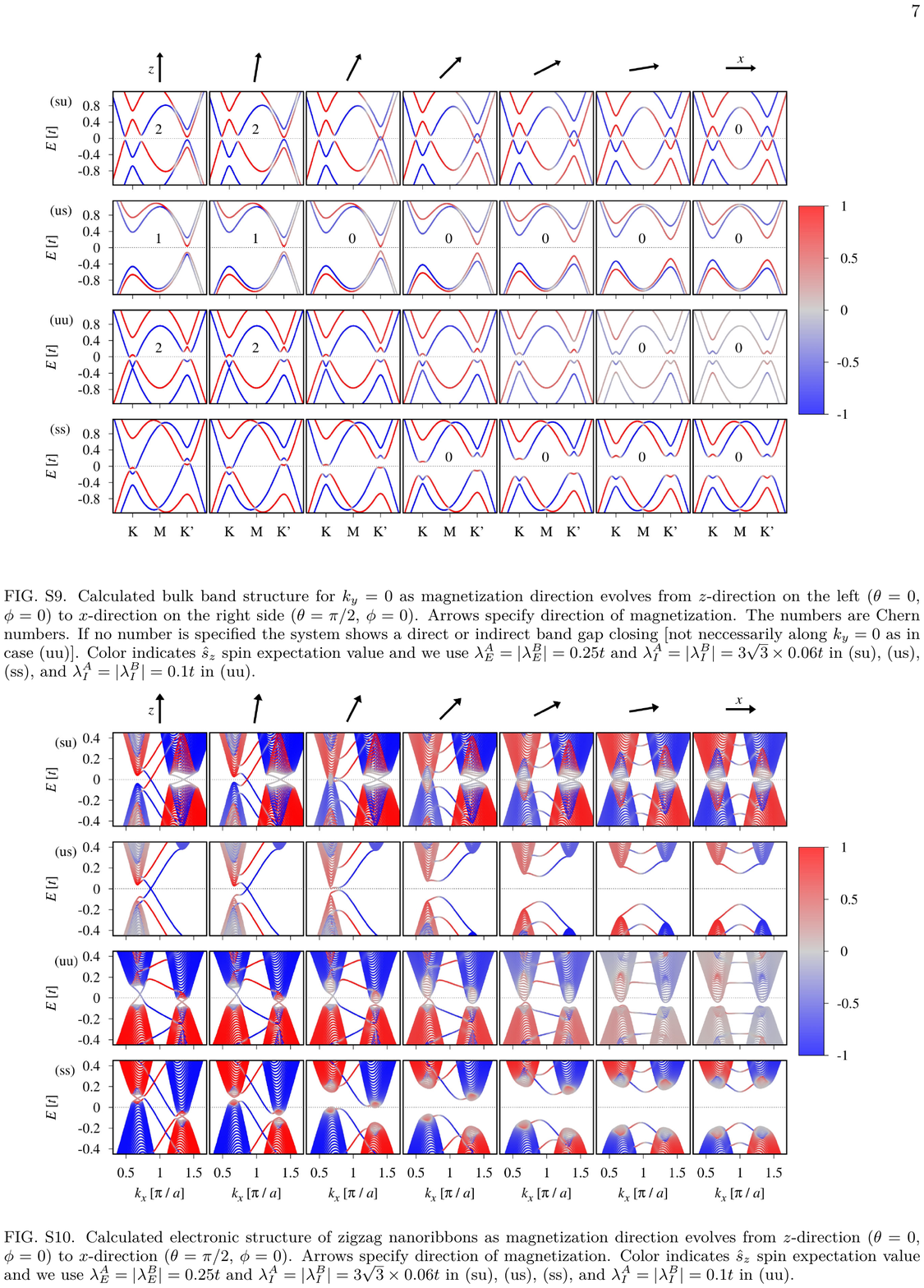}}
\end{figure*}
\begin{figure*}[htb]
	{\includegraphics[width=2.1\columnwidth]{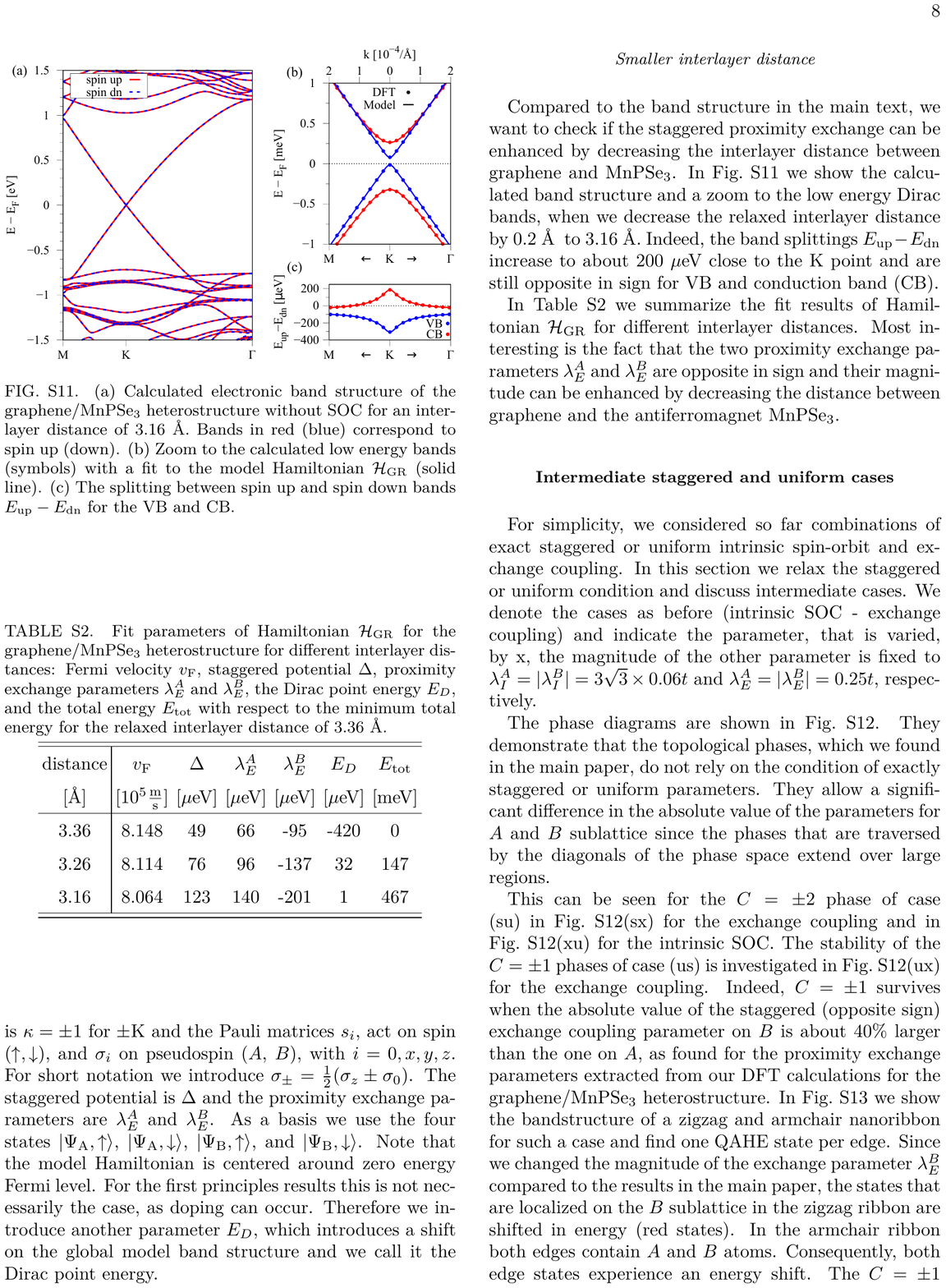}}
\end{figure*}
\begin{figure*}[htb]
	{\includegraphics[width=2.1\columnwidth]{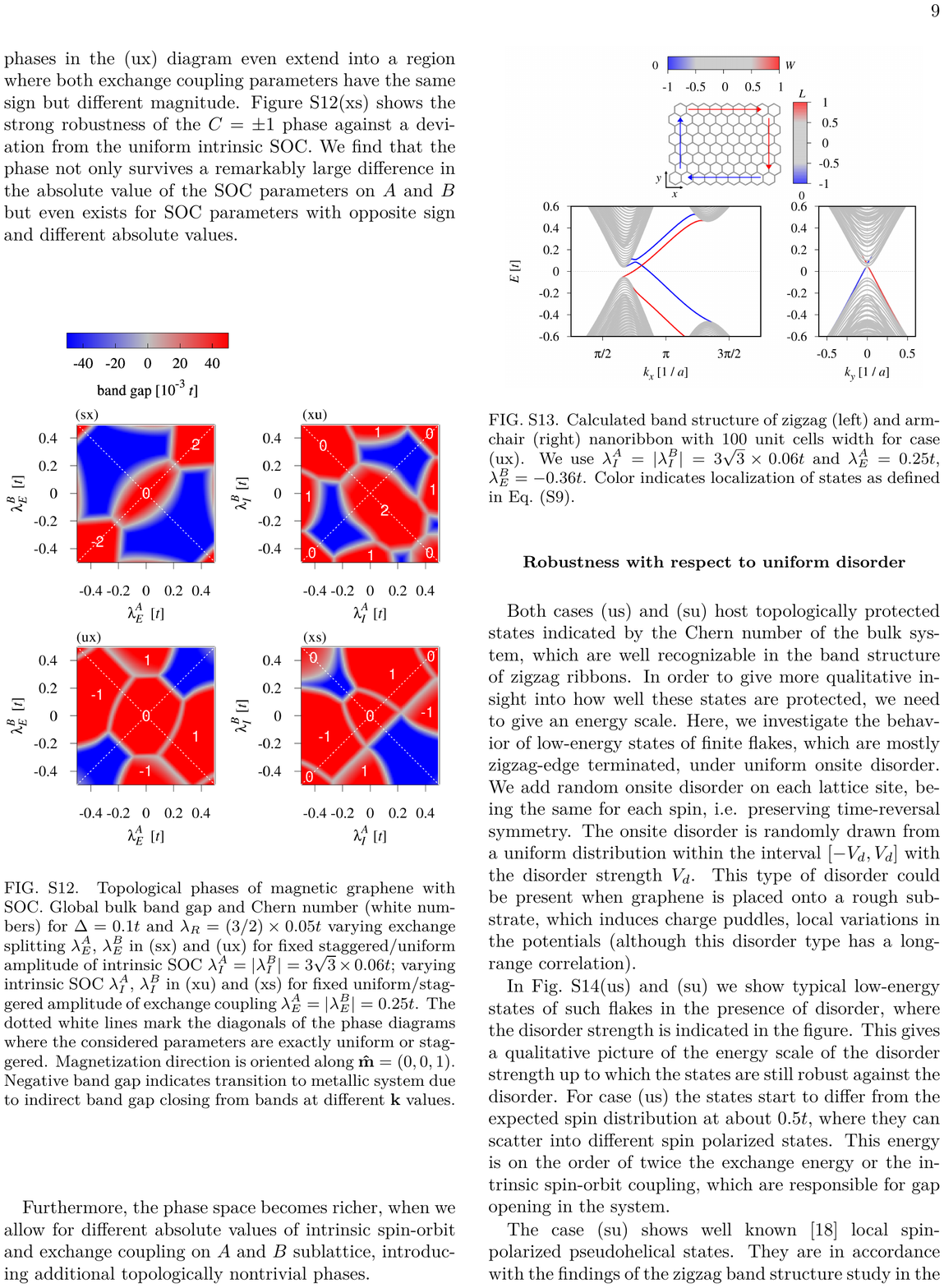}}
\end{figure*}
\begin{figure*}[htb]
	{\includegraphics[width=2.1\columnwidth]{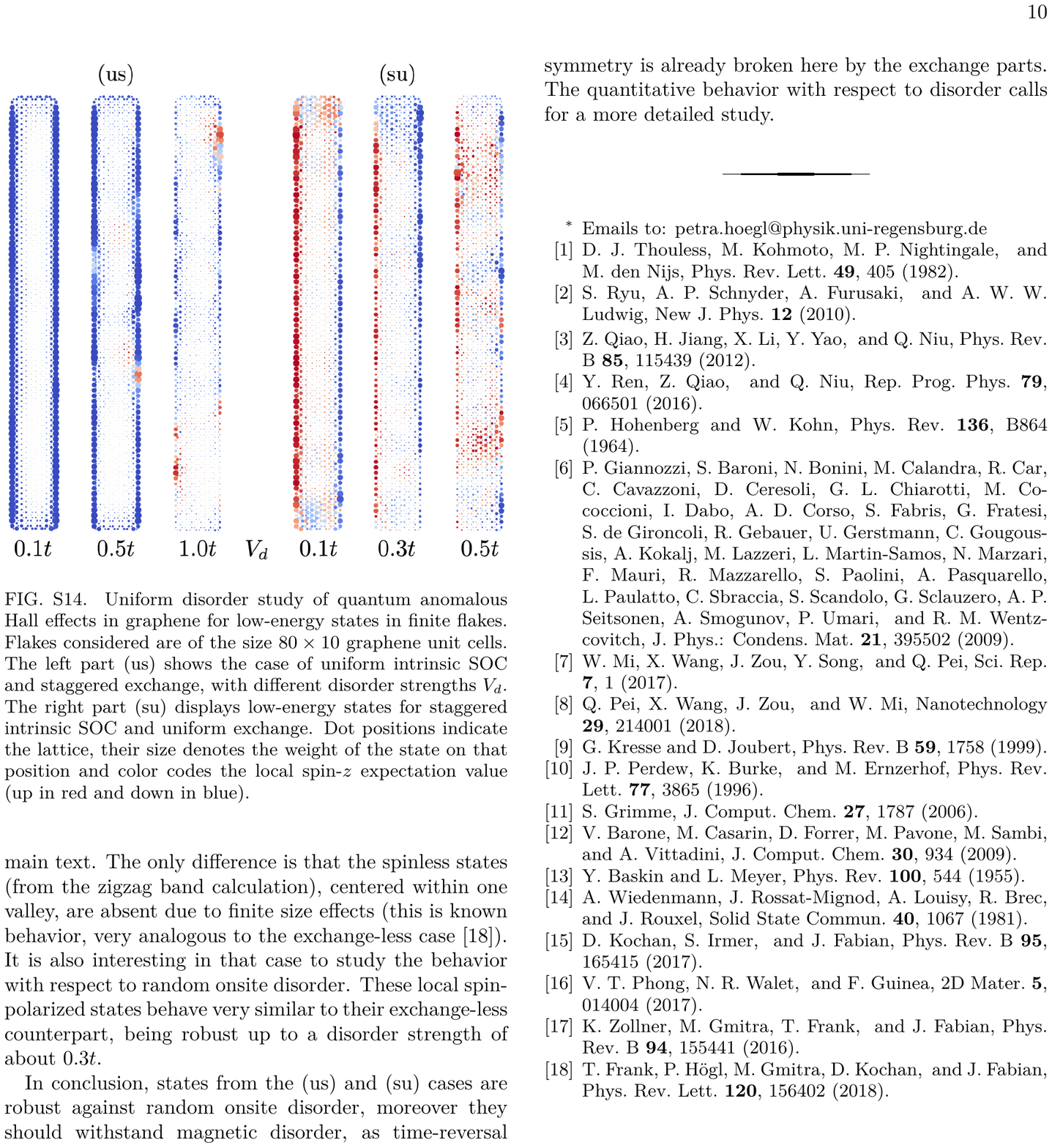}}
\end{figure*}
	
\end{document}